\documentclass[
	a4paper, 
	10pt, 
	unnumberedsections, 
	twoside, 
]{LTJournalArticle}

\usepackage{listings}

\runninghead{} 

\footertext{\textit{RISC-V Summit Europe, Paris, 12-15th May 2025}} 

\title{Programming RISC-V accelerators via Fortran} 

\author{
	Nick Brown\textsuperscript{1}\thanks{Corresponding author: \href{mailto:n.brown@epcc.ed.ac.uk}{\tt Nick Brown (n.brown@epcc.ed.ac.uk)}}, Jake Davies\textsuperscript{1}, Felix LeClair\textsuperscript{2}
}

\date{\footnotesize\textsuperscript{\textbf{1}}EPCC, The University of Edinburgh, 47 Potterrow, Edinburgh, UK \\ \textsuperscript{\textbf{2}}Tenstorrent, 2600 Great America Way, Santa Clara, CA, USA}

\renewcommand{\maketitlehookd}{%
	\begin{abstract}
A range of RISC-V based accelerators are available and coming to market, and there is strong potential for these to be used for High Performance Computing (HPC) workloads. However, such accelerators tend to provide bespoke programming models and APIs that require codes to be rewritten. In scientific computing, where many of the simulation code are highly complex, extensive, and written in Fortran, this is not realistic. In this extended abstract we present an approach that enables driving such architectures via Fortran, avoiding code redevelopment.
	\end{abstract}
}

\begin{document}

\maketitle 


\section{Introduction}

Whilst RISC-V has grown rapidly in areas such as embedded computing, it is yet to gain significant traction in High Performance Computing (HPC). However, as we move further into the exascale era the HPC community will be faced by a range of new challenges, for instance the requirement to decarbonise their workloads, and there is the potential for RISC-V to play an important role.

Arguably, it is likely that we will first see adoption of RISC-V in HPC via PCIe accelerator cards. These can can be easily fitted into existing systems and because it enables centres to \emph{dip their toe} into the RISC-V ecosystem it limits their risk as other parts of the supercomputer remain the same. Indeed, there are a range of RISC-V PCIe accelerators cards that are shipping, such as Esperanto's ET-SoC and the Tensix family from Tenstorrent, with other products such as Inspire Semiconductor's Thunderbird having been announced. However, the major challenge associated with all of these is that to actually run codes on them then the developer must learn a new programming model, restructure their codes to map to the architecture, and leverage the vendor API.

Fortran is the lingua franca of scientific computing, indeed around 65\% of codes running on ARCHER2, the UK national supercomputer, are written in Fortran and these account for around 70\% of the machine's runtime. Ultimately, developers of these high performance codes want to run more complex problems at reduced time to solution, and the specialisation provided by RISC-V accelerators means that they can potentially provide this whilst also delivering energy benefits. However, a major challenge to adoption of such technologies is the requirement for scientific programmers to significant restructure their codes, potentially also having to rewrite them in different programming languages.

\subsection{MLIR}
Since it was first merged into mainstream LLVM in 2019, MLIR has become a popular for developing compilers. Comprising Intermediate Representation (IR) dialects, and transformations which undertake optimisations and convert the IR between dialects, it is possible to mix dialects which are at different levels of abstraction and progressively lower between them. Ultimately, MLIR provides reuse of compiler infrastructure, and via the MLIR framework one can define their own IR dialects and transformations. 

\subsection{Flang}
Flang is the LLVM community's Fortran compiler and leverages MLIR by providing it's own Fortran IR (FIR) and High Level Fortran IR (HLFIR) dialects. However, only a subset of MLIR standard dialects are integrated with Flang, and Flang itself transforms straight from HLFIR+FIR into LLVM-IR without using any of the existing MLIR transformations or optimisations.

Conversely, the \emph{mlir-opt} MLIR driver tool is unaware of the Flang dialects and it is not possible to drive the wide range of MLIR transformations and optimisations via Flang's IR. To this end in \ref{brown2024fully} we developed a transformation pass that lowers Flang's HLFIR and FIR dialects into standard MLIR dialects. The first benefit of this is that the user is then able to leverage the existing MLIR transformations which are developed and maintained by a large community, including many vendor, to generate LLVM-IR. The second benefit is that it provides a much wider range of potential target architectures including GPUs.

\section{Flang for RISC-V accelerators}

\begin{figure*}[htb]
\centering
 \includegraphics[width=0.85\textwidth]{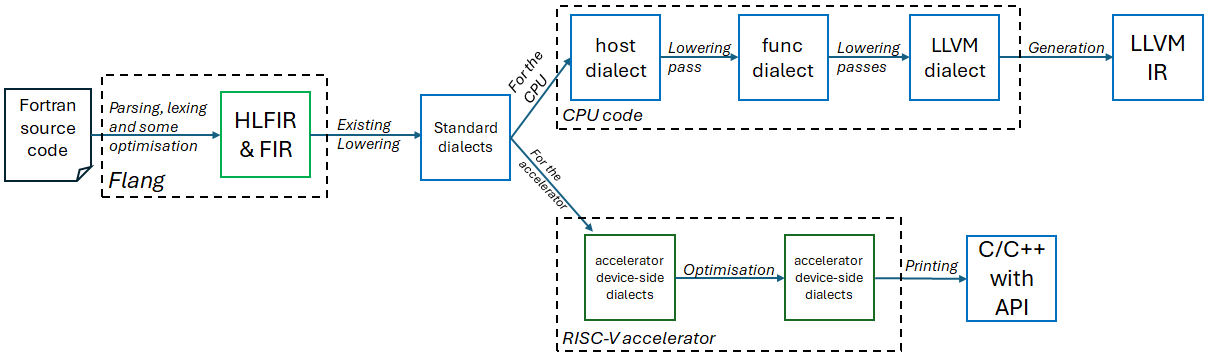}
\caption{Illustration of our approach lowering Flang to target RISC-V accelerators}	
\label{fig:flang-rv}
\end{figure*}

Figure \ref{fig:flang-rv} provides an overview of our approach,  where the existing work of \cite{brown2024fully} lowers Fortran into the standard dialects and a transformation is provided which lowers into the specific host and device dialects for the RISC-V accelerator. Some PCIe based RISC-V accelerators already provide an MLIR-based compiler stack and-so we can then leverage their existing dialects in combination with their compilation pipeline to generate the resulting executables. 

However, many of these accelerators either do not provide an MLIR stack or such a stack is immature. In such a case, as per Figure \ref{fig:flang-rv}, we develop a backend for these accelerators which comprises host and device-side dialects that map one-to-one to the accelerator API. A lowering is then developed that converts the host-side dialect to the \emph{func} dialect, calling runtime functions and eventually into LLVM-IR. On the device-side we develop a \emph{printer} which accepts the device specific dialects and standard MLIR dialects, such as \emph{memref} for memory management, and this prints out target code comprising a programming language, commonly C or C++, calling into the device's API.

\subsection{Tenstorrent example}

Tenstorrent ship RISC-V PCIe accelerator cards that are built upon their Tensix technology. Each Tensix core comprises five RISC-V cores; one for data movement in, one for data movement out, and three drive a 16384 wide vector unit. The Wormhole n300, for example, contains 128 Tensix cores. The decoupling of data movement from compute makes this a very interesting potential architecture for HPC, and indeed early experiments porting a scientific computing workload to the Grayskull delivered similar performance to a 24-core Xeon Platinum CPU but at five times less energy usage \cite{brown2024accelerating}. However, to run codes on this architecture programmers must learn a new architecture and significantly recast their applications. 

We developed a Tenstorrent specific backend which comprises a host dialect and three device-side dialects, one for data movement, one for circular buffers between RISC-V cores, and one for compute. We also developed a printer that, from the device-side dialects, generates C++ that calls into the Metalium API.

\begin{lstlisting}[language=fortran, frame=lines, label=lst:example_ftn, numbers=none, caption=Example Fortran code running Single-precision A times X Plus Y (SAXPY) on the Tenstorrent accelerator card (argument declarations omitted for brevity)]
subroutine saxpy(a, x, y, n)
  ...
  !$omp omp target parallel &
  !$omp& do simd num_threads(20) simdlen(32)
  do i=1, n
    y(i) = a * x(i) + y(i)
  end do
  !$omp end target parallel do simd  
end subroutine
\end{lstlisting}

A question is how, in Figure \ref{fig:flang-rv}, to lower from the standard dialects into the device-specific ones that map to the RISC-V accelerator. The programmer drives this via OpenMP target offload, and Listing \ref{lst:example_ftn} illustrates Single-precision A times X Plus Y (SAXPY) written in Fortran and offloaded to the Tenstorrent PCIe accelerator via OpenMP. HPC programmers are already familiar with OpenMP, both for threaded and GPU programming so it is a natural choice. The code example in Listing \ref{lst:example_ftn} will run the loop in parallel over two Tensix cores, due to the \emph{num\_teams(2)}, leveraging the SIMD capabilities of each Tensix core. 

\section{Conclusions}

We have described offloading Fortran code to RISC-V based accelerators via Flang. OpenMP provides a clear abstraction which can be used to drive such an offloading, and MLIR is a powerful compiler technology for supporting these accelerators because it enables the sharing of compiler infrastructure between them.





\printbibliography 


\end{document}